\documentclass{aa}  
\usepackage[varg]{txfonts}
\usepackage{rotating}
\usepackage{ulem}
\usepackage{xcolor}
\usepackage[breaklinks, colorlinks, citecolor=blue, urlcolor = blue]{hyperref}      
\usepackage{csquotes}
\usepackage{units}
\usepackage{mathtools}
\usepackage{float}
\usepackage[caption = false]{subfig}
\usepackage{graphicx}
\usepackage{amsmath}
\usepackage{footmisc}
\usepackage{multirow}

\bibpunct{(}{)}{;}{a}{}{,}

\def\gsim{ \lower .75ex \hbox{$\sim$} \llap{\raise .27ex \hbox{$>$}} }
\def\lsim{ \lower .75ex\hbox{$\sim$} \llap{\raise .27ex \hbox{$<$}} }
\def\sw{{\it Swift}}
\def\fe{{\it Fermi}}

\def\G{$\Gamma_{0}$}

\begin{document}

\title{GRB~190114C: from prompt to afterglow?}
\titlerunning{$\gamma$--ray spectral evolution of 190114C}
\authorrunning{M. E. Ravasio}
\author{ M. E. Ravasio\inst{1,2}, G. Oganesyan \inst{3,4}, O. S. Salafia \inst{1,2}, G. Ghirlanda\inst{1,2},
G. Ghisellini\inst{1}, M. Branchesi\inst{3,4}, S. Campana\inst{1}, S. Covino\inst{1}, R. Salvaterra\inst{5}\\ }  

\institute{$^1$ INAF -- Osservatorio Astronomico di Brera, via E. Bianchi 46, I-23807 Merate, Italy.\\
$^2$ Dipartimento di Fisica G. Occhialini, Univ. di Milano Bicocca, Piazza della Scienza 3, I-20126 Milano, Italy.\\
$^3$ Gran Sasso Science Institute, Viale F. Crispi 7, I-67100, L’Aquila, Italy \\
$^4$ INFN - Laboratori Nazionali del Gran Sasso, I-67100, L’Aquila, Italy\\
$^5$ INAF - Istituto di Astrofisica Spaziale e Fisica Cosmica , via E. Bassini 15, I-20133 Milano, Italy
}

\date{}

\abstract{GRB 190114C is the first gamma-ray burst detected at very high energies (VHE, i.e., $>300$ GeV) by the MAGIC Cherenkov telescope. 
The analysis of the emission detected by the \textit{Fermi} satellite at lower energies, in the 10 keV -- 100 GeV energy range, up to $\sim$50 s (i.e.,~before the MAGIC detection) can hold valuable information. We analyze the spectral evolution of the emission of GRB 190114C as detected by the \textit{Fermi} Gamma-Ray Burst Monitor (GBM) in the 10 keV -- 40 MeV energy range up to $\sim$60 s. 
The first 4 s of the burst feature a typical prompt emission spectrum, which can be fit by a smoothly broken power-law function with typical parameters. 
Starting on $\sim$4 s post-trigger, we find an additional nonthermal component that can be fit by a power law. This component rises and decays quickly. 
The 10 keV -- 40 MeV flux of the power-law component peaks at $\sim$6 s; it reaches a value of 1.7$\rm \times 10^{-5} erg\, cm^{-2}\, s^{-1}$. 
The time of the peak coincides with the emission peak detected by the Large Area Telescope (LAT) on board {\it Fermi}. The power-law spectral slope that we find in the GBM data is remarkably similar to that of the LAT spectrum, and the GBM+LAT spectral energy distribution seems to be consistent with a single component.
This suggests that the LAT emission and the power-law component that we find in the GBM data belong to the same emission component, which we interpret as due to the afterglow of the burst.
The onset time allows us to estimate that the initial jet bulk Lorentz factor \G\, 
is about 500, depending on the assumed circum-burst density.
}

\maketitle

\section{Introduction}

Soon after its launch, the {\it Fermi} satellite has been detecting\footnote{\url{http://fermi.gsfc.nasa.gov/ssc/observations/types/grbs/lat_grbs/table.php}.} $\text{about}$ 14 gamma-ray bursts (GRBs) per year on average with its Large Area Telescope (LAT) in the high-energy (HE) range between a few MeV to 100 GeV  \citep{Ackermann2013}. The
{\it Fermi}/LAT GRBs  confirm the detections by the Astro Rivelatore Gamma ad Immagini Leggero (Agile/GRID -- \citet{Giuliani2008,Giuliani2010,Delmonte2011}) and the earlier results of the {\it Compton Gamma Ray Observatory}/EGRET \citep{Sommer1994,Hurley1994,Gonzalez2003}.
Until very recently, observations of GRBs emission at very high energies (VHE) by Imaging Atmospheric Cherenkov Telescopes (IACT) resulted only in upper limits (\citet{Aliu2014}, \citet{Carosi2015,Hoischen2017}). GRB 190114C is the first burst detected at $>300$ GeV by the Major Atmospheric Gamma Imaging Cherenkov Telescopes (MAGIC) \citep{Mirzoyan2019GCN}. 

Gammy-ray burst emission in the 100 MeV -- 100 GeV energy range as detected by LAT typically starts with a short delay with respect to the trigger time of the keV--MeV component \citep{Omodei2009,Ghisellini2010,Ghirlanda2010} and extends until after the prompt emission. This behavior has also been observed in short GRBs \citep{Ghirlanda2010,Akermann2010}. While the early HE emission (simultaneous with the keV--MeV component) shows some variability, its long-lasting tail decays smoothly. A possible transition from an early steep decay ($\propto t^{-1.5}$) to a shallower regime ($\propto t^{-1}$) has been reported \citep{Ghisellini2010,Ackermann2013} and a faster temporal decay in brighter bursts has been claimed \citep{Panaitescu2017}. 

During the prompt emission phase (as detected, e.g., by~the Gamma Ray Burst Monitor, GBM, on board the \textit{Fermi} satellite), the LAT spectrum can either be the extension above 100 MeV of the typical sub-MeV GRB spectrum (which is usually fitted with the Band function;  \citet{Band1993}), or it requires an additional spectral component in the form of a power law (PL), as in GRB 080916C, 110713A \citep{Ackermann2013}, 090926A \citep{Yassine2017}, and 130427A  \citep{Ackermann2014}. In a few bursts, this additional PL component has been found to extend to the X-ray range ($<20$ keV;  e.g., 090510,  \citet{Akermann2010}, and 090902B,  \citet{Abdo2009}). 
When the prompt emission has ceased, the LAT spectrum is
often fit by a PL with photon index  $\Gamma_{\rm PL}\sim -2$.

The interpretation of the HE emission of GRBs is still debated (see \citet{Nava2018} for a review). It has been proposed that the LAT emission that extends after the end of the prompt emission is the afterglow that is produced in the external shock
that is driven by the jet into the circum-burst medium \citep{Kumar2009, Ghisellini2010, Kumar2010}. The mechanism that causes this might be synchrotron emission. The correlation of the LAT luminosity with the prompt emission energy \citep{Nava2014} and the direct modeling of the broadband spectral energy distribution (initially in a few bursts, \citet{Kumar2009,Kumar2010} and then in a larger sample \citet{Beniamini2015}) support the hypothesis of a synchrotron origin.

A possible problem with the synchrotron interpretation are VHE photons (tens of GeV), which exceed the theoretical limit of synchrotron emission from shock-accelerated electrons. This limit is $\sim$70 MeV in the comoving frame (\citet{Guilbert1983}, see also \citet{deJaeger1996} and \citet{Lyutikov2010} for a lower value of about 30 MeV), but downstream magnetic field stratification \citep{Kumar2012} or acceleration in magnetic reconnection layers \citep{Uzdensky2011,Cerutti2013} can alleviate this apparent discrepancy. 

The deceleration of the jet by the interstellar medium is expected to produce a peak 
in the afterglow light curve at a time $t_{\rm p}$
that corresponds to the transition from the coasting to the deceleration phase \citep{Sari1999}. 
$t_{\rm p}$ depends on the blast wave kinetic energy $E_{\rm k}$, on the density of the 
circum-burst medium (and its radial profile),
and on the initial bulk Lorentz factor $\Gamma_{0}$ 
(representing the maximum velocity that the jet attained, i.e.,~that of the coasting phase). 
Therefore, by deducing $E_{\rm K}$ from the prompt emission and making an assumption on the circum-burst medium density, it is possible to estimate $\Gamma_{0}$   \citep{Molinari2007,Ghirlanda2012,Ghirlanda2018} for large samples of GRBs.

If the GeV component is afterglow produced by the external shock, the time 
$t_{\rm p}$ provides an estimate of $\Gamma_{0}$ (see also \citet{Nava2017}), as shown for the first time in the case of the LAT-detected GRB 090510 \citep{Ghirlanda2010}. 
The shorter $t_{\rm p}$, the larger $\Gamma_0$: LAT bursts have the shortest times $t_{\rm p}$ \citep{Ghirlanda2018} and therefore provide the highest values of $\Gamma_0$ up to $\sim$1200 (GRB 090510 -- \citet{Ghirlanda2018}). As discussed in \cite{Ghisellini2010}, this might indicate that a large $\Gamma_0$ helps to accelerate very high energy electrons, which emit at high photon energies. 
Furthermore, even a small fraction of photons of the prompt phase can be scattered by the circum-burst medium and act as targets for the $\gamma$--$\gamma \to$ $e^\pm$ process: this enhances the lepton abundance of the medium, thus making shock acceleration of the leptons more efficient \citep{Beloborodov2005, Ghisellini2010}.

While the LAT emission, which in some cases is detected up to hours after the end of the prompt, seems to be of external origin, a possible challenge is the interpretation of the early LAT emission that is detected during the prompt phase. 
It has been argued  \citep{zhang2011,He2011} that the very early LAT emission has 
an internal origin \citep{Bosnjak2009} because it can be due to inverse Compton-scattered synchrotron photons of the prompt (SSC). 
The delay of the GeV emission as measured by LAT could be explained by inverse Compton emission that occurred in the Klein--Nishina regime at early times \citep{Daigne2012,Bosnjak2009}.  While recent findings seem to support a synchrotron origin of keV--MeV photons \citep{Oganesyan2017,Oganesyam2018, Ravasio2018}, the presence of a soft excess ($<$50 keV) that is clearly detected so far in GRB 090902B \citep{Abdo2009}, GRB 090510 \citep{Ackermann2010}, and GRB 090926A \citep{Yassine2017}, represents a challenge for the SSC interpretation (but see \citet{Toma2011}) and would be more easily interpreted as the the low-energy extension of the GeV afterglow component.

This paper is based on the study of the emission of GRB 190114C (\S2) as detected by the 
GBM in the 10 keV -- 40 MeV energy range, up to 61 s after the trigger. We also consider data from the Burst Alert Telescope (BAT) and the X-Ray Telescope (XRT) on board the {\it Neil Gehrels Swift Observatory} in three time intervals.
While the properties of GRB 190114C are similar to other bursts detected by LAT,  emission that might extend up to the TeV energy range as detected by MAGIC \citep{Mirzoyan2019GCN} makes this event unique so far.
Data extraction and analysis are presented in \S3 and in \S4, where we show the appearance and temporal evolution of a nonthermal power-law spectral component  
starting from 4 s after the trigger. 
In \S 5 we discuss our results and their implications.

\section{GRB 190114C}

On 14 January 2019 at 20:57:03 UT, both the \textit{Fermi}/GBM and the \textit{Swift}/BAT were triggered by GRB 190114C \citep{Hamburg2019GCN,Gropp2019GCN}. 
The burst was also detected in hard X-rays by the SPI-ACS instrument on board \textit{INTEGRAL}, with evidence for long-lasting emission \citep{Minaev2019GCN}, 
by the Mini-CALorimeter (MCAL) instrument on board the {\it AGILE} satellite \citep{Ursi2019GCN}, by the Hard X-ray Modulation Telescope 
(HXMT) instrument on board the \textit{Insight} satellite \citep{Xiao2019GCN}, and by
Konus-Wind \citep{Frederiks2019GCN}.

Remarkably, this burst was the first to be detected at very high energies by a Cherenkov telescope: MAGIC was able to point the source 50 s after the {\it Swift} trigger, revealing the burst with a significance $>20\sigma$ at energies $>$300 GeV \citep{Mirzoyan2019GCN}. The burst was also detected by LAT. It 
remained in its field of view until 150 s after the GBM trigger \citep{Kocevski2019GCN}.

The redshift was first measured by the Nordic Optical Telescope (NOT) \citep{Selsing2019GCN} (soon confirmed by the Gran Telescopio Canarias GTC, \citet{Castro-Tirado2019GCN}), with the value $z = 0.4245 \pm 0.0005$.

The fluence (integrated in the 10--1000 keV energy range) measured by the GBM is $3.99\times 10^{-4}\pm 8\times 10^{-7}$ erg cm$^{-2}$ and the peak photon flux (with 1 s binning in the same energy range) is $246.86 \pm 0.86$ cm$^{-2}$ s$^{-1}$  \citep{Hamburg2019GCN}. As reported in \citet{Hamburg2019GCN}, the corresponding isotropic equivalent energy and luminosity are   $E_\mathrm{iso}\sim 3\times 10^{53}\,\mathrm{erg}$ and $L_\mathrm{iso}\sim 1\times 10^{53}\,\mathrm{erg}\,s^{-1}$, respectively. 
These values make this burst consistent with the $E_{\rm peak}$--$E_{\rm iso}$
\citep{Amati2002}
and $E_{\rm peak}$--$L_{\rm iso}$ \citep{Yonetoku2004} correlations
\citep{Frederiks2019GCN}.

The prompt emission of GRB 190114C is characterized by a first (multi-peaked) 
pulse that lasted $\sim$ 5.5 s, followed by a second weaker and softer pulse from
15 to 22 s after trigger (as shown in the top panel of Fig. \ref{lc}), and then a weaker and long tail that lasted up to some hundreds of seconds \citep{Hamburg2019GCN,Minaev2019GCN}.

\section{Data analysis}
\subsection{Fermi/GBM}
The GBM is composed of 12 sodium iodide (NaI, 8 keV--1 MeV) and 2 bismuth germanate
(BGO, 200 keV--40 MeV) scintillation detectors \citep{Meegan2009}. 
We analyzed the data of the three brightest NaI detectors with a viewing angle smaller than $60^{\circ}$ (n3, n4, and n7) and both the BGO detectors (b0 and b1). 
In particular, we selected the energy channels in the range 8--900\,keV for NaI detectors, excluding the channels in the range 25--40 keV because of the iodine K--edge at 33.17 keV\footnote{\url{https://fermi.gsfc.nasa.gov/ssc/data/analysis/GBM\_caveats.html}} and 0.3--40\,MeV for BGO detectors.
Spectral data files and the corresponding response matrix files (\texttt{.rsp2}) were obtained from the online 
archive\footnote{\url{https://heasarc.gsfc.nasa.gov/W3Browse/fermi/fermigbrst.html}} , and the spectral analysis was performed with the public software {\sc rmfit}-~(v.~4.3.2).
To model the background, we selected background spectra in time intervals well before and after the burst (approximately -130 : -10 s and 210 : 370 s from the trigger time) and modeled them with a polynomial function up to the third order.
We used time-tagged event (TTE) data, and rebinned them with a time resolution of 0.3 s during the first emission episode of the burst. After the first emission episode, we rebinned the data in progressively longer time bins up to the second minor peak of the light curve (from  $\sim 15$ s to $\sim 23$ s), which was analyzed as a single bin. Finally, we analyzed the 23--61 s time interval as two consecutive time bins
(23--47 s and 47--61 s).

\subsection{Swift: BAT and XRT  data}
We also considered BAT data extracted for three time bins, 6--6.3 s, 47--61 s, and 87--232 s, both as a check of the consistency with the parameters of the fit obtained in the same time intervals from GBM data and as a way to extend our analysis to later times. We downloaded BAT event files from the \sw\ data archive\footnote{\url{http://heasarc.gsfc.nasa.gov/cgi-bin/W3Browse/swift.pl}}. 
To extract BAT spectra, we used the latest version of the {\sc heasoft} package (v6.25). We generated BAT spectral files with the {\tt batbinevt} task, applying the correction for systematic errors with the {\tt batupdatephakw} and {\tt batphasyserr} tasks. We generated response files with the {\tt batdrmgen} tool. We adopted the latest calibration files (CALDB release 2017--10--16). 

In addition, we retrieved XRT event files from the \sw/XRT archive\footnote{\url{http://www.swift.ac.uk/archive/}}. The source and background files were extracted with  the {\tt xselect} tool. We removed the central region of the XRT image to avoid pile-up effects, following the procedure described in \cite{Romano_06}. 
We generated an ancillary response file with the {\tt xrtmkarf} task. We excluded all the channels below 1.5 \,keV because an apparent low-energy excess has been reported in \cite{xrtreport}. We then rebinned the energy channels using the the {\it grppha} tool, requiring at least 40 counts per bin. 

We used the multiplicative XSPEC models {\tt tbabs} and {\tt ztbabs} to account for Galactic and intrinsic absorption of the X-ray spectrum by neutral hydrogen \citep{wilms00}. The value of Galactic neutral hydrogen column density in the direction of GRB 190114C was found from \cite{Kalberia_05}. The intrinsic column density $\rm 7.7\times 10^{22} cm^{-2}$ was estimated by fitting  the late-time X-ray spectrum ($5.6 \times 10^4 - 5.7 \times 10^{5}$ s). 

\subsection{Fitting models}
A preliminary analysis of the GBM spectrum was reported in \cite{Hamburg2019GCN}: the time-integrated spectrum from 0 to $38.59$ s (which includes the two pulses of the burst but also the inter-pulse interval) was fit with a Band function, finding $E_{\rm peak} = 998.6 \pm 11.9$ keV, $\alpha = -1.058 \pm 0.003$, and $\beta = -3.18 \pm 0.07$. 
In addition, the authors also reported a strong statistical preference for an extra power-law component.

In our time-resolved analysis, we fit the spectra with a smoothly broken power-law (SBPL, see \citet{Ravasio2018} for a description of the functional form). The SBPL is one of the empirical functions that is generally used to model GRB spectra \citep{Kaneko2006,Gruber2014}.
The SBPL is made of two power laws, with spectral indices $\alpha$ and $\beta$, which are smoothly connected at the break energy (usually corresponding to the $\nu F_\nu$ peak of the spectrum, $E_{\rm peak}$). As in \citet{Ravasio2018}, the curvature parameter was kept fixed at n = 2.

Because an additional power-law component was reported in \cite{Hamburg2019GCN}, we also added an additional power-law component in the fitting procedure, with two free parameters, the normalization $N$ and the spectral index $\Gamma_{\rm PL}$.

\section{Results}

\begin{figure*}[htb]
\centering
\hskip -1.5 cm %%for printer version
\centering\includegraphics[scale=0.65]{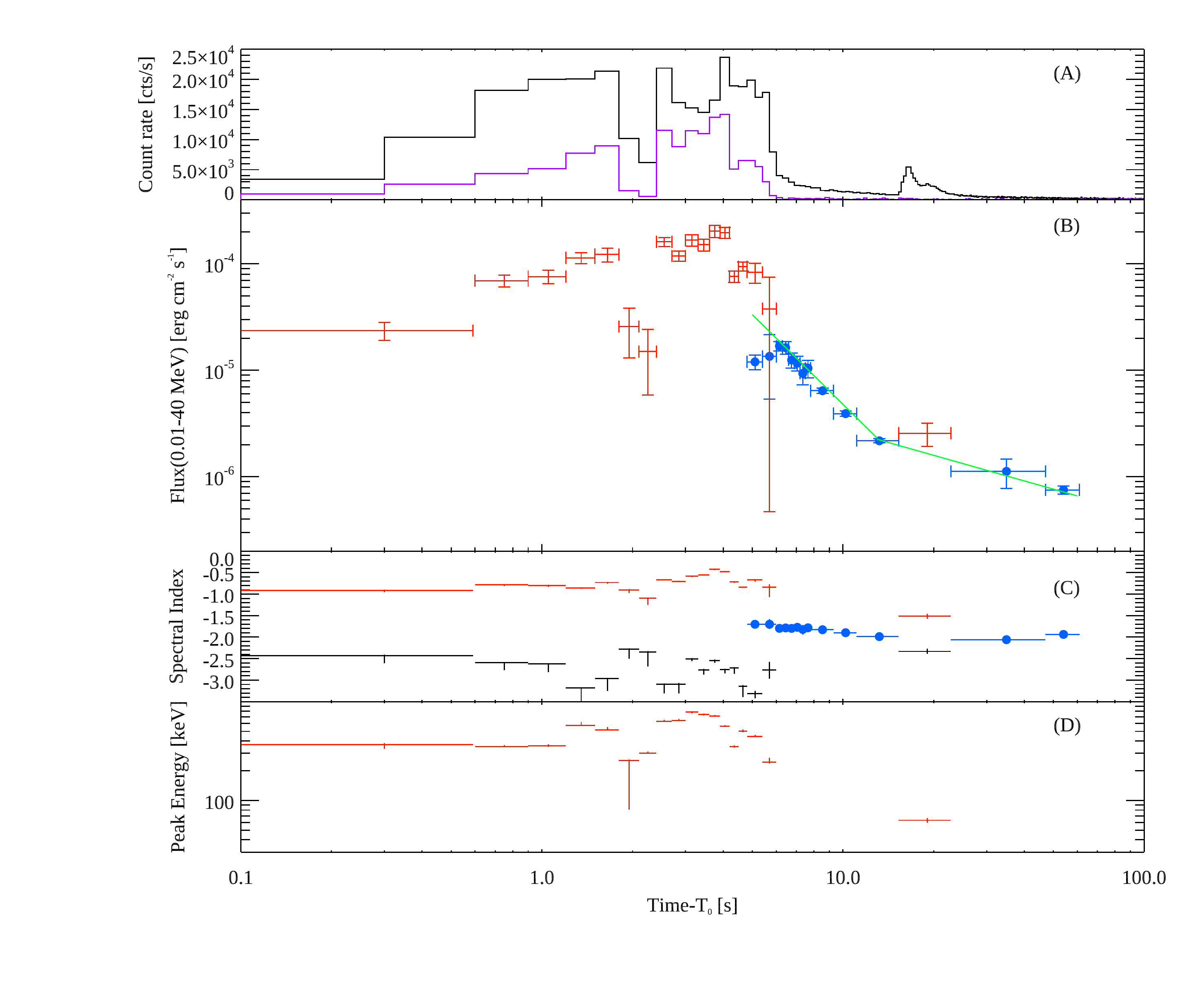}
\vskip -1 cm
\caption{Spectral evolution of GRB190114C. Two spectral components are shown: 
Smoothly broken power law (SBPL, red symbols) and power law (PL, blue circles). 
$1\sigma$ errors are shown. Panel A: Count rate light curve (black solid line for GBM NaI detector 3 and purple solid line for GBM BGO detector 0).
Panel B: Flux (integrated in the 10 keV -- 40 MeV energy range) of the two spectral components. The green line is a power law with slope --2.8 up to 15 s, with slope --1 when the decay of the flux is shallower. 
Panel C shows the temporal evolution of the spectral photon index of the SBPL (red and black symbols) and of the PL (blue symbols). 
Panel D shows the evolution of the peak energy ($E_{peak}$) of the SBPL model.}
\label{lc}
\end{figure*}

Fig.~\ref{lc} shows the results of the time-resolved spectral analysis of GBM data. We find that all spectra belonging to the first emission episode  (from 0 s to 4.8 s) are  reasonably well fit by an SBPL model and no additional power-law component is required. The low- and high-energy spectral indices of the SBPL model are shown in panel (C) of Fig.~\ref{lc} (red and black symbols, respectively). 
Their values are consistent with the typical distributions obtained from the analysis of large samples of GBM bursts (\citet{Goldstein12,Gruber2014, Nava2011,Kaneko2006}). 
The peak energy (panel D in Fig.~\ref{lc}) evolves and tracks the flux of the light curve, with an average value of $E_{\rm peak} = 510 \pm 170$\,keV.

The additional power-law component starts in the 4.8--5.4 s and 5.4--6.0 s time bins, where the superposition of an SBPL and a PL component is preferred over the SBPL component alone (an F-test yields a 6 and 7.5 $\sigma$ preference for the SBPL+PL model in the first and second bin, respectively).

The power-law component reaches its peak in the time bin 6--6.3 s, with a flux of $1.7\, \pm 0.2 \times 10^{-5}\, \mathrm{erg\, cm^{-2}\, s^{-1}}$, integrated in the energy range 10 keV--40 MeV.
From 6.3 s onward, the spectrum is well fit (p-value $>0.3$ in all bins) by a single power-law PL component, with no increase in the goodness of fit when the SBPL component is added. Moreover, when we tried to fit with the SBPL function, the peak energy $E_{\rm peak}$ was completely unconstrained, and the values found for the two spectral indices $\alpha$ and $\beta$ are consistent with each other within the errors.
The single power-law spectral slope is shown by the blue symbols in panel (C) of Fig.~\ref{lc}. Its 10 keV -- 40 MeV flux is shown by the blue symbols in panel (B).

The average spectral slope of the PL component in the time interval 4.8--15.3 s is $\Gamma_{\rm PL} = -1.81 \pm 0.08$, similar to the spectral slope found in the LAT data (at >100 MeV, \citet{Kocevski2019GCN}) in the same time interval \citep{Wang2019}. After $\sim$10 s, the slope of the power law becomes constant and settles at the --2 value, again similar to the LAT index.
The second emission episode was fit by an SBPL, 
with $\alpha = -1.51 \pm 0.06$, $\beta = -2.33 \pm 0.06,$ and $E_{\rm peak} = 63 \pm 3$ keV. 
The parameters of the additional power law were not constrained, and the fit did not improve when it was included.
After 22.8 s, the spectrum was again well fit by a power law alone, with index 
$\Gamma_{\rm PL}\sim -2$.
The flux of the PL component (panel B of Fig.~\ref{lc}) decayed steeply from the peak up to 15 s (a reference green line $\propto t^{-2.8}$ is shown). 
From 15--50 seconds, the temporal decay of the flux was consistent with $t^{-1.0}$. 

\begin{figure*}[h]
% \centering
%\hskip -1 cm
%\vskip -5 cm %%for referee version
\vskip -7.5 cm %%for printer version
\includegraphics[width=\textwidth]{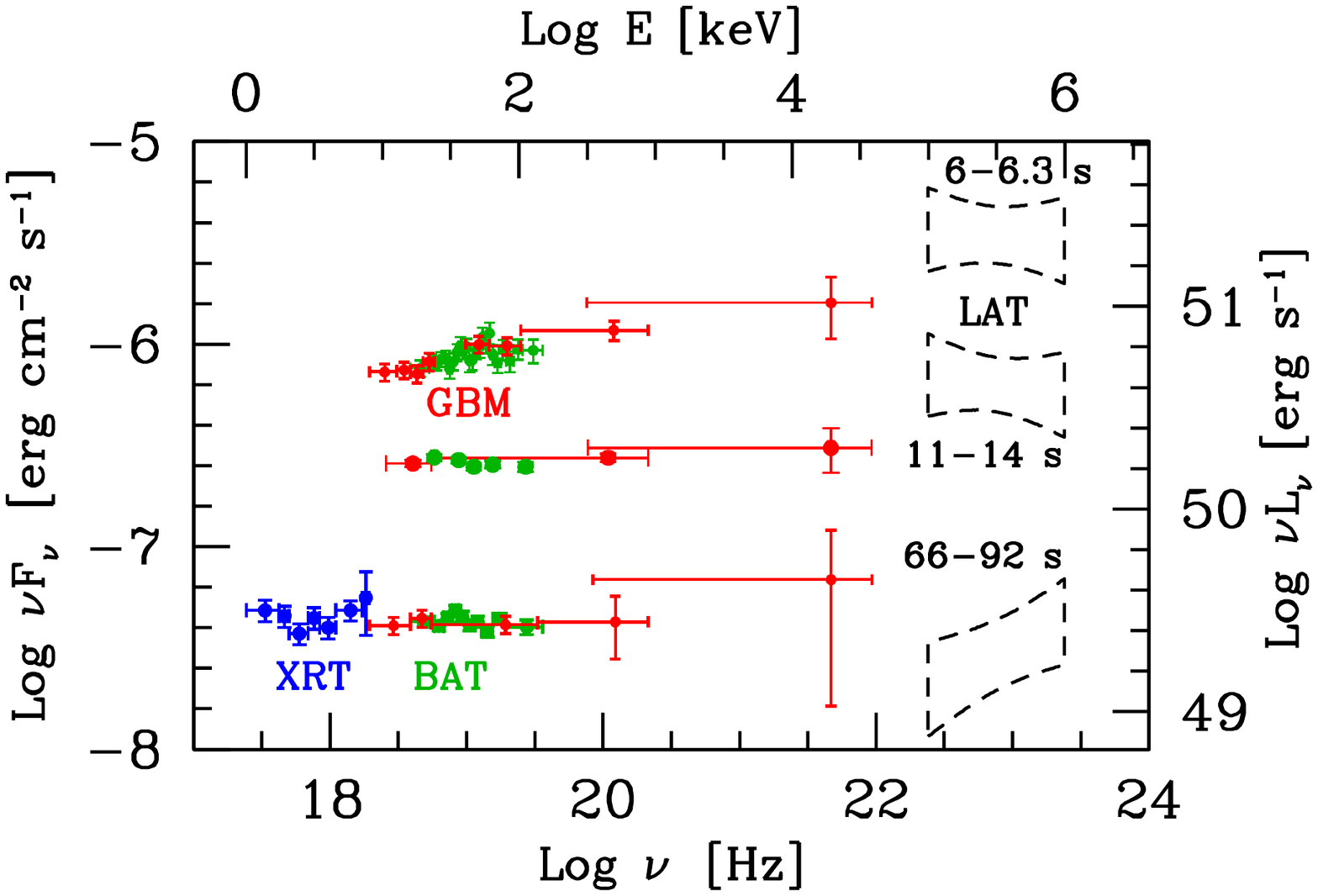}
%\vskip -5 cm %%for referee version
\vskip -6 cm %%for printer version
\caption{ X--ray to GeV SED of GRB 190114C at three specific times:
at 6-6.3 s, when the power-law component peaks in the GBM data (see panel (B) of Fig.\ref{lc}, blue symbols), at 11--14 s, and at 66--92 s (as labeled). We show the GBM, BAT, and XRT data (the latter deabsorbed, as described in the text). Errors and upper limits on the data points represent 1$\sigma$. The LAT butterflies represent the range of fluxes and indices of the power law reported in the analysis of \citet{Wang2019}.}
\label{fig:sed}
\end{figure*}

We also added BAT data for the time intervals 6.0--6.3 s and 11--14 s. In both time bins, BAT+GBM data were fit together with a single PL, from which we obtained best-fit parameters that were consistent with the analysis of GBM data alone. We also verified that BAT data alone for the first time bin result in power-law parameters that were fully consistent with those derived from the fit of the GBM spectrum alone.
Fig. \ref{fig:sed} shows the spectral energy distribution of the three time intervals (as labeled). Spectral data used in the fits are BAT+GBM for interval 6--6.3 s and 11--14 s. XRT+BAT+GBM spectra are shown for the last time bin (66--92 s). \citet{Wang2019}  analyzed the LAT spectrum of GRB 190114C by fitting the high-energy data with a power-law model. Fig. \ref{fig:sed} also shows the LAT flux and spectral index with butterflies (including the corresponding uncertainties) for the same time intervals, to be compared with our results.

The GBM and BAT data appear to be connect to the LAT emission, as analyzed by \citet{Wang2019}. In the two time intervals 6--6.3 s and 11--14 s, the photon indices of the LAT spectrum are $\Gamma_{\rm PL} = -2.06 \pm 0.30$ and $\Gamma_{\rm PL} = -2.10 \pm 0.31,$ respectively, which are consistent with the values we obtained from our analysis. The LAT emission is slightly higher than the GBM extrapolation (by less than 60\%: less than 2$\sigma$).
Moreover, we analyzed XRT+BAT+GBM data from 66 s to 92 s to check again for consistency with the LAT flux given in \citet{Wang2019} and also to track the power-law evolution at later times. As shown in Fig.~\ref{fig:sed}, the LAT flux is still consistent with extrapolation of the joint XRT+BAT+GBM data fit.
From our analysis, the fit of XRT+BAT+GBM data from 66 s to 92 s with a PL function results in a spectral slope $\Gamma_{\rm PL} = -2.01 \pm 0.05$, which is only marginally consistent with the values obtained by \citet{Wang2019} for the LAT data ($\Gamma_{\rm PL} = -1.67 \pm 0.27$). 
We note, however, that the spectral slopes reported in \citet{Wang2019} have large uncertainties and show a rapid variability.
In summary, Fig. \ref{fig:sed} shows that the keV-MeV and GeV emissions have a similar time decay and similar slopes, suggesting that they belong to the same component. However, because of the uncertainties on the LAT spectral parameters reported in \citet{Wang2019}, the possibility that the GeV and keV-MeV data belong to two different components cannot be excluded.

%-----------------------------------------------------

%-----------------------------------------------------
%-----------------------------------------------------

\subsection{Estimate of \G}
Several slightly different formulae can be used to derive the bulk Lorentz factor $\Gamma_0$
of the coasting phase from the observational data.
The required parameters are
i) the peak time of the light curve $t_{\rm p}$;
ii) the isotropic equivalent kinetic energy of the jet $E_{\rm K}$ after the emission of the prompt radiation;
iii) the circum-burst density $n$, which is responsible for the deceleration of the jet, and 
iv) its radial profile.

Usually, it is assumed that the observed isotropic equivalent energy radiated in the prompt phase $E_{\rm iso}$ is a fraction $\eta$ of the kinetic energy, implying $E_{\rm K} = E_{\rm iso}/\eta$, typically with $\eta=0.1$ or 0.2.
The density is assumed to have a radial profile $n\propto R^{-s}$ ($R$ is the distance from the central engine originating the GRB).
We considered the case of a uniform density ($s=0$), 
or a steady stellar wind density profile ($s=2$). In the latter case, the density depends on the mass rate $\dot M_{\rm w}$ of the wind and its velocity $v_{\rm w}$ \citep{Chevalier2000},  $n(R) = \dot M_{\rm w}/(4\pi v_{\rm w} R^2 m_\mathrm{p})$.

The different formulae used to calculate $\Gamma_0$ have been thoroughly discussed in \cite{Ghirlanda2018}. 
As in that paper, we used the formula derived in \cite{Nava2013}
\begin{equation} 
\Gamma_{0} 
= \left[\frac{(17-4s)(9-2s)3^{2-s}}{2^{10-2s}\pi(4-s)}
\left(\frac{E_\mathrm{K}}{n_0 m_{\rm p}c^{5-s}}\right)\right]^{\frac{1}{8-2s}}
t_{\rm p, z}^{-\frac{3-s}{8-2s}}
\label{n13}
,\end{equation}
which for the two different cases of homogeneous medium (s=0) and wind density profile (s=2), becomes
\begin{equation} 
\Gamma_{0} \propto \left(\frac{E_\mathrm{iso}}{\eta n_0 m_{\rm p}c^{5}}\right)^{\frac{1}{8}}
t_{\rm p, z}^{-\frac{3}{8}}                        \qquad  (s=0)\\
\label{n14}
\end{equation}
\begin{equation} 
\Gamma_{0} \propto \left(\frac{E_\mathrm{iso}}{\eta n_0 m_{\rm p}c^{3}}\right)^{\frac{1}{4}}
t_{\rm p, z}^{-\frac{1}{4}}                        \qquad  (s=2)
\label{n15}
.\end{equation}

Here $t_{\rm p}$ is measured in the source cosmological rest frame, that is, $t_{p,z} = t_{\rm p}/(1+z)$, $m_{\rm p}$ is the mass of the proton, and $n_0$ is the normalization of the circum-burst density profile, that is,~$n(R)=n_0\,R^{-s}$.

Assuming $E_{\rm iso} = 2.6\times 10^{53}$ erg calculated from 0 to 6 s, 
$\eta=0.2$, $t_{\rm p}=6$ s,
through Eq. \ref{n13} we estimate
$\Gamma_{0}\sim 700 \pm 26$ (520 $\pm$ 20) in the case of a homogeneous medium with density $n=1$\, cm$^{-3}$ ($n=10$ cm$^{-3}$ ). 
For a wind medium with $\dot M_{\rm w} =10^{-5} M_\odot/{\rm yr}$ and  $v_{\rm w}=10^3$ km/s ($v_{\rm w}=10^2$ km/s), following the relation $n_0 = \dot M_{\rm w} / 4\pi v_{\rm w} m_p $, the initial bulk Lorentz factor is $\Gamma_{0}\sim 230 \pm 6$ (130 $\pm$ 3).
The errors are only statistical and were calculated using the uncertainties on the observables $E_{\rm iso}$ and  $t_{\rm p}$; the errors do not include the unknown uncertainties on parameters $\eta$ and $n_0$ .

Table 2 in \cite{Ghirlanda2018} lists the coefficients that are required to calculate $\Gamma_0$
for all the other proposed formulae for the homogeneous and for the wind case.
The resulting $\Gamma_0$ values 
differ
at most by a factor of 2.
The computed values are similar to those found for other GBRs detected by LAT, which show a peak in the light curve in the LAT energy band \citep{Ghirlanda2018}.

\section{Discussion}

Our results indicate that a power-law component appears at $\sim 4$ s after trigger in the GBM data, that it peaks at 6 s, and then declines.
This temporal behavior matches that of the flux above 100 MeV, as seen by the LAT.
Fig.~\ref{fig:sed} shows that the emission in the two detectors (GBM and LAT) joins smoothly, with a consistent slope (within the errors).
It is therefore compelling to interpret the two power laws seen in LAT and GBM as belonging to a single emission component.
We propose that this nonthermal emission is produced by the external shock that is driven by the jet into the circum-burst medium.  Its peak marks the jet deceleration time, that is,~onset time of the afterglow. 

The reasons leading to this interpretation are
i) they appear after the trigger of the prompt event, and peak when most of the 
prompt emission energy has already been radiated;
ii) they last much longer than the prompt emission;
iii) they are characterized by a spectral index ($\Gamma_{\rm PL}\sim -2$) typical 
of the known afterglows;
iv) with the exception of the early variable phases, their light curve smoothly decays with a temporal slope typical of the known afterglows.

We remark that this is not the first time that a power law is detected in the hard X-rays in addition to the spectral components that are usually seen during the prompt emission phase. 
A component like this was well visible in GRB 090202B, another burst that was very strong in the LAT band (\citet{Rao2013} and references above). The observation of the onset of the afterglow in the hard X-ray band is new, however, as is that it was found to be simultaneous within the uncertainties with the peak of the LAT light curve.
This is especially important in this burst because of the MAGIC detection.

Our results imply that emission in the energy range between 10 keV and 30 GeV is produced by a single mechanism.
If this mechanism is synchrotron or inverse Compton emission, this in turn implies that the energy of the underlying electron distribution must extend over more than three orders of magnitude.

We also know that the MAGIC telescope revealed photons above 300 GeV \citep{Mirzoyan2019GCN} despite the strong absorption due to the extragalactic optical-infrared background \cite[e.g.,][]{Franceschini2008} that is expected for $z=0.425$.
If the maximum synchrotron energy is 
$h\nu_{\rm max}= m_{\rm e} c^2 /\alpha_{\rm F}\sim$ 70 MeV in the comoving frame, as theoretically predicted in the case of shock acceleration \citep{Guilbert1983, deJaeger1996}, then the radiation above 300 GeV might be interpreted as due to another process, most likely inverse Compton or synchrotron self-Compton emission.
On the other hand, the observed maximum photon energy detected by LAT, 22.9 GeV 15 s after trigger, does not violate the comoving 70 MeV limit if the bulk Lorentz factor $\Gamma$ at this time is higher than 450. 
For this value to be consistent with $\Gamma_0$, that is,~the bulk Lorentz of the jet before it starts to be decelerated by the circum-burst medium, (assuming a prompt efficiency $\eta=0.2$) the circum-burst medium must not be too dense, with a number density $n\lesssim 30\,\mathrm{cm^{-3}}$ in the homogeneous case, or the progenitor stellar wind to be slightly faster and/or less massive than usually assumed, to satisfy $\dot M_\mathrm{w,-5}\, v_\mathrm{w,8} \lesssim 0.02$ (where $\dot M_\mathrm{w,-5} = \dot M_\mathrm{w}/(10^{-5}\,\mathrm{M_\odot\,yr^{-1}})$ and $v_\mathrm{w,8}=v_\mathrm{w}/(10^{8}\,\mathrm{cm\,s^{-1}}$)).

Alternatively, the entire spectral energy distribution from the keV to the TeV energy range
could be inverse Compton emission, possibly by Compton scattering 
off IR--optical radiation.
In this case, the MAGIC emission should connect smoothly
with the LAT spectrum (i.e., it should not be harder). 
Therefore the MAGIC flux and spectrum will give crucial information about the
origin of the entire high-energy spectrum of GRBs.

\section*{Acknowledgments}
We would like to thank Lara Nava for fruitful discussions. 
M. E. R. is grateful to the Observatory of Brera for the kind hospitality.
This research has made use of data obtained through the High Energy Astrophysics Science Archive Research Center Online Service, provided by the NASA/Goddard Space Flight Center, and specifically, this work made use of public \fe-GBM data. We acknowledge INAF-Prin 2017 (1.05.01.88.06) for support and the Italian Ministry for University and Research grant "FIGARO" 1.05.06.13. We also would like to thank for the support of the implementing agreement ASI-INAF n.2017-14-H.0.

\bibliographystyle{aa} 
\bibliography{references}

\end{document}